\documentclass[journal,twoside]{IEEEtran}

\ifCLASSINFOpdf
\else
\fi
\usepackage{multirow}
\usepackage{graphicx}
\usepackage{epstopdf}
\usepackage{stfloats}
\usepackage{subfigure}
\usepackage{microtype}
\usepackage{setspace}
\usepackage{cite}
\usepackage{color}
\usepackage{amsmath}
\usepackage{array}
\usepackage{pifont}
\usepackage{color}

\usepackage{amsmath,amssymb,amsfonts}
\usepackage{algorithmic}
\usepackage{textcomp}

\begin{document}

\title{Geometry-Based Stochastic Line-of-Sight Probability Model for A2G Channels under Urban Scenarios}

\author{Qiuming~Zhu,~\IEEEmembership{Member,~IEEE,}
        Fei~Bai, Minghui~Pang, Jie~Li,
        Weizhi~Zhong,~\IEEEmembership{Member,~IEEE,}
        Xiaomin~Chen,~\IEEEmembership{Member,~IEEE,}
        Kai~Mao  
\thanks{This work was supported by the Fundamental Research Funds for the Central Universities (No.~NS2020026, No.~NS2020063),
the National Key Scientific Instrument and Equipment Development Project (No.~61827801), Aeronautical Science Foundation of China (No.~201901052001),
and Open Foundation for Graduate Innovation of NUAA (No.~KFJJ20200416). \emph{(Corresponding author: Qiuming~Zhu and Jie~Li)}}
\thanks{Q.~Zhu is with the Key Laboratory of Dynamic Cognitive System of Electromagnetic Spectrum Space, College of Electronic and Information Engineering, Nanjing University of Aeronautics and Astronautics, Nanjing 211106, China. He is also with State Key Laboratory of Integrated Services Networks, Xidian University, Xian 710000, China (e-mail: zhuqiuming@nuaa.edu.cn).}
\thanks{F.~Bai, M.~Pang, J.~Li, X.~Chen and K.~Mao are with the Key Laboratory of Dynamic Cognitive System of Electromagnetic Spectrum Space, College of Electronic and Information Engineering, Nanjing University of Aeronautics and Astronautics, Nanjing 211106, China (e-mail: \{baifei, pangminghui, lijie{\_}evelyn, chenxm402, maokai\}@nuaa.edu.cn).}
\thanks{W.~Zhong is with the Key Laboratory of Dynamic Cognitive System of Electromagnetic Spectrum Space, College of Astronautics, Nanjing University of Aeronautics and Astronautics, Nanjing 211106, China (e-mail: zhongwz@nuaa.edu.cn).}}

\markboth{IEEE Transactions on Antennas \& Propagation, VOL. XX, NO. XX, 2021}
{Geometry-Based Stochastic Line-of-Sight Probability Model for A2G Channels under Urban Scenarios}
%



\maketitle

\begin{abstract}
Line-of-sight (LoS) path is essential for the reliability of air-to-ground (A2G) communications, but the existence of LoS path is difficult to predict due to random obstacles on the ground. Based on the statistical geographic information and Fresnel clearance zone, a general stochastic LoS probability model for three-dimensional (3D) A2G channels under urban scenarios is developed. By considering the factors, i.e., building height distribution, building width, building space, carrier frequency, and transceiver's heights, the proposed model is suitable for different frequencies and altitudes. Moreover, in order to get a closed-form expression and reduce the computational complexity, an approximate parametric model is also built with the machine-learning (ML) method to estimate model parameters. The simulation results show that the proposed model has good consistency with existing models at the low altitude. When the altitude increases, it has better performance by comparing with that of the ray-tracing Monte-Carlo simulation data. The analytical results of proposed model are helpful for the channel modeling and performance analysis such as cell coverage, outage probability, and bit error rate in A2G communications.
\end{abstract}

\begin{IEEEkeywords}
A2G channels, stochastic LoS probability, Fresnel zone, machine-learning, ray-tracing.
\end{IEEEkeywords}

%
\IEEEpeerreviewmaketitle

\section{Introduction}
%
%
%
%
\IEEEPARstart{U}{nmanned} aerial vehicles (UAVs) have shown great growth in various fields due to their high mobility, low cost, and easy deployment \cite{Xiao20_ITJ, Liu20_CC, Mao20_Sensors}. The air-to-ground (A2G) communication technology is promising for the sixth generation (B5G/6G) mobile communication systems \cite{WCX21_SCIC, Wang20_VTM, Li19_ITJ, Zhu20_IWCMC, Rinaldi21_TB}. Note that A2G in this paper includes the UAV, airship, air balloon, and other aircrafts to the ground. The A2G channel has exhibited different characteristics from land mobile channels, i.e., 3D scattering environments and valid scatterers around the ground station \cite{Khawaja19_CST, Zhu18_trans}. Especially, the line-of-sight (LoS) path dominates the reliability of A2G link, while it's time-varying and difficult to predict due to random scatterers \cite{YinXF21_TAP, Zhu19_IET}. Therefore, it is vital to explore and build a LoS probability model for channel modeling and performance evaluation \cite{Roy19_TAP, Zhong19_CC}.
\begin{figure*}[!b]
	\centering
	\includegraphics[width=150mm]{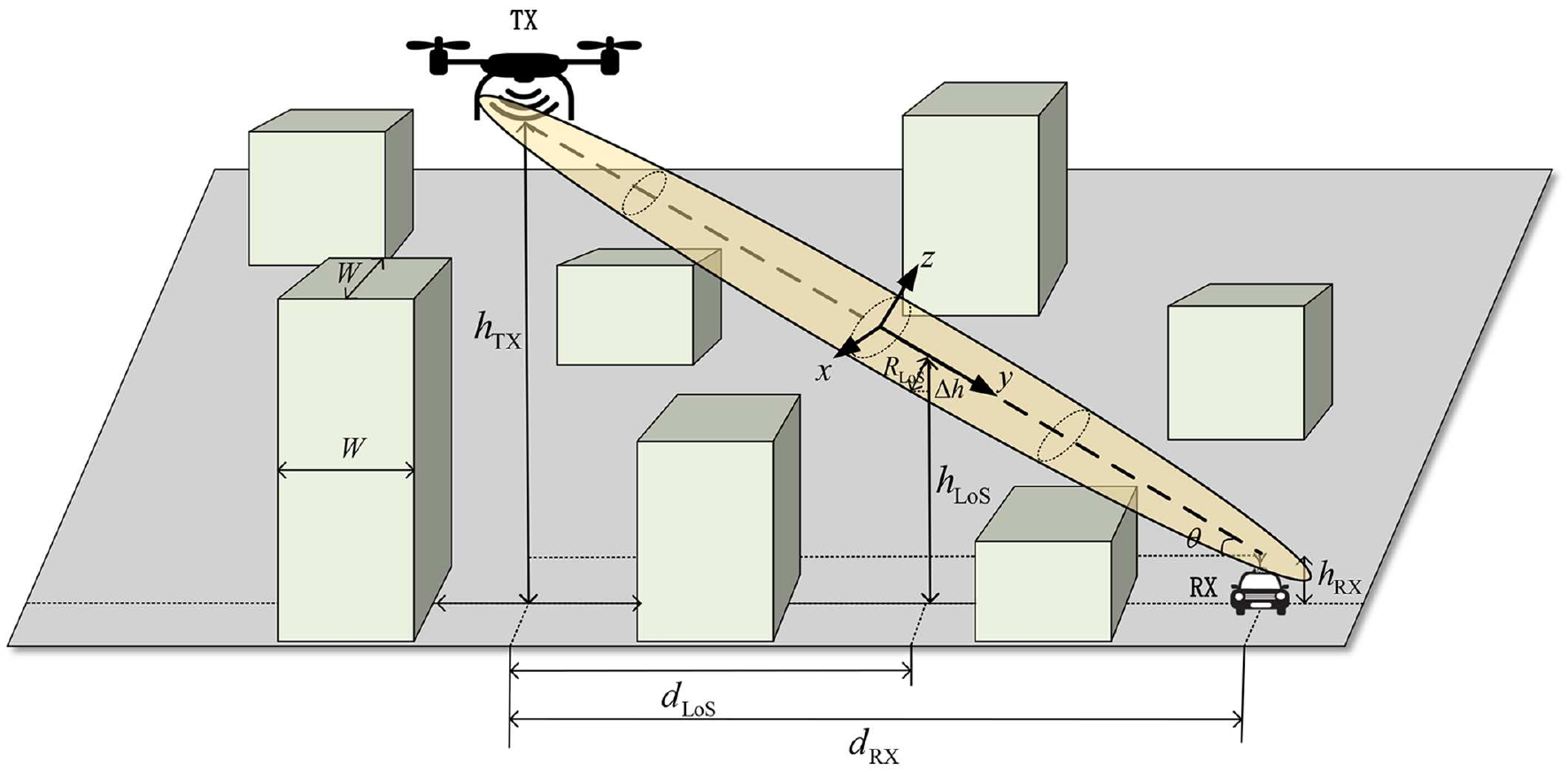}
	\caption{LoS propagation under urban scenarios.}
    \label{fig:1}
\end{figure*}
\par There are limited theoretical or measurement studies in the literature for LoS probability \cite{3GPP, ITU-R2135, WINNER, 5GCM, Samimi15_WCL, Lee18_TAP, Jarvelainen16_WCL, ITU-R1410, Holis08_TAP, Hourani14_WCL, Hourani20_WCL, Liu18_CL, Cui20_ITJ}. Several standardized channel models, e.g., the Third Generation Partnership Project (3GPP) TR 38.901 \cite{3GPP}, International Telecommunication Union-Radio (ITU-R) M.2135-1 \cite{ITU-R2135}, WINNER II \cite{WINNER}, and 5G Channel Model (5GCM) \cite{5GCM}, have their own LoS probability prediction methods. However, most of them are based on the measurement results and only suitable for the low-altitude (or land mobile communication) scenarios. These empirical methods can only be accurate for some specific environments, but not applicable for general environments with different deployments of buildings. Recently, some researchers studied the LoS probability by deterministic methods, e.g., the ray-tracing (RT) method \cite{Samimi15_WCL, Lee18_TAP}, and the point cloud method \cite{Jarvelainen16_WCL}. Whereas, this kind of methods requires an accurate digital map with detailed material information of undergoing environment, which is unpractical in most applications.
\par On the other hand, the analytical method based on the statistical geographic information can provide acceptable precision without loss of simplicity and generality \cite{ITU-R1410, Holis08_TAP, Hourani14_WCL, Hourani20_WCL, Liu18_CL, Cui20_ITJ}. It described the built-up scenarios and analyzed the LoS probability in a stochastic way. For example, a widespread analytical method proposed in the ITU-R Rec. P.1410 model \cite{ITU-R1410}, used three parameters to describe the geometrical statistics of urban areas. The distribution of building centers was uniform and the distribution of building heights followed Rayleigh distribution. Note that the building width factor was not included in \cite{ITU-R1410}. Inspired by this idea, several stochastic LoS probability models for urban scenarios were addressed in \cite{Holis08_TAP, Hourani14_WCL, Hourani20_WCL, Liu18_CL, Cui20_ITJ}. For example, by conducting massive RT simulations for all possible locations under statistical urban areas, the LoS probability was fitted by a function with respect to the elevation angle in \cite{Holis08_TAP, Hourani14_WCL}. However, these two models were only appropriate for high-altitude case. The authors in \cite{Hourani20_WCL} described the buildings by cylinders with random heights. By analyzing the data of Melbourne city, it recommended that the building height obeyed the log-normal distribution and the building centers distribution followed the Poisson point process. Furthermore, the authors in \cite{Liu18_CL, Cui20_ITJ} took the frequency factor into account by considering the Fresnel zone. This idea was more reasonable for the radio wave propagation by checking the LoS component with the Fresnel clearance zone. However, both of the aforementioned two methods have high computational complexity. Thus, this paper aims to find a general LoS probability model with a good tradeoff between complexity and accuracy. The main contributions and innovations are summarized as follows:
\par 1) Considering the statistical geographic information and Fresnel zone, a general stochastic model is proposed to predict the blockage probability of radio wave LoS propagation under urban scenarios. The proposed model which considers the geometric factors such as the locations of terminals, building height distribution, building width, and building space, is applicable for different altitudes. Besides, by taking the Fresnel ellipse radius into account, the factor of carrier frequency is introduced into the model.
\par 2) The effects of building width and frequency on the LoS probability are investigated based on the proposed analytical model. Moreover, an approximate model with back propagation neural network (BPNN)-based framework for parameter generation is also developed, which can be used to derive the closed-form solutions of channel model and system performance. This proposed approximate model can greatly reduce the computational complexity with good accuracy.
\par 3) Massive numerical simulations with RT-based Monte-Carlo method are conducted to demonstrate and evaluate the performance of the proposed stochastic LoS probability model. Different virtual urban scenes are reconstructed based on environment-dependent parameters, and are applied to obtain the average LoS probability by RT simulations. The simulation results show that our proposed model has better performance at high altitude case compared with the existing models.
\par The remainder of this paper is organized as follows. Section II gives the new definition of LoS probability from the perspective of radio wave propagation. In Section III, the analytical stochastic model is derived and a corresponding approximate parametric model based on machine-learning (ML) is also proposed for typical urban scenes. The RT-based Monte-Carlo simulation results and validation are given in Section IV. Finally, conclusions are drawn in Section V.
\setcounter{equation}{6}
\begin{figure*}[!b]
\begin{equation}
{{P}_{\text{LoS}}}({{d}_{\text{RX}}},{{h}_{\text{TX}}},{{h}_{\text{RX}}},\psi )=\prod\limits_{i=\text{1}}^{{{N}_{b}}}{{{P}_{i}}}=\prod\limits_{i=\text{1}}^{{{N}_{b}}}{\left[ 1-\exp \left( -\dfrac{{{\left[ {{h}_{\text{TX}}}-\left( \dfrac{i-0.5}{{{N}_{b}}} \right)({{h}_{\text{TX}}}-{{h}_{\text{RX}}}) \right]}^{2}}}{2{{\gamma }^{\text{2}}}} \right) \right]}
\label{7}
\end{equation}
\end{figure*}
\section{LoS Propagation Path}
\par The urban scenario is the most typical application area for the future A2G communications, where randomly distributed buildings are referred as the main obstacles. A typical A2G communication link under the urban scenario is shown in Fig.~1. In order to make the LoS probability prediction more general, the environment-dependent geometric information can be described in a stochastic way. Note that the distributions of building layout and building height have been considered for LoS probability prediction in \cite{ITU-R1410, Hourani14_WCL}, but the building width factor is ignored for simplicity \cite{Holis08_TAP, Hourani20_WCL}. In the Fig.~1, ${{h}_{\text{RX}}}$, ${{h}_{\text{TX}}}$ represent the heights of UAV as the transmitter (TX) and vehicle as the receiver (RX), respectively, ${{d}_{\text{RX}}}$ is the horizontal distance between the TX and the RX, $W$ represents the mean width of buildings and $\theta$ is the elevation angle.
\par In \cite{ITU-R1410}, LoS propagation, which is independent with the carrier frequency, is regarded as the geometric LoS path (or the optical direct path). When the positions of TX and RX are known, the LoS path means that the straight line between the TX and RX is not blocked by any buildings. On the contrary, if one or more buildings obstruct the straight line, it is regarded as the NLoS propagation. Based on this assumption, the LoS probability can be defined as
\setcounter{equation}{0}
\begin{equation}	
{{P}_{\text{LoS}}}=\prod\limits_{i=\text{0}}^{{{N}_{b}}}{{{P}_{i}}=}\prod\limits_{i=\text{0}}^{{{N}_{b}}}{P({{h}_{i}}<{{h}_{_{\text{LoS}}}})}
\label{1}
\end{equation}
\noindent where ${{h}_{i}}$ is the building height, ${{h}_{\text{LoS}}}$ is the maximum allowed height of buildings, and ${{N}_{b}}$ is the total number of buildings along the LoS path. The NLoS probability can be easily obtained by ${{P}_{\text{NLoS}}}=1-{{P}_{\text{LoS}}}$.
\par Note that the above definition of LoS path is not accurate from the perspective of radio wave propagation, since the radio radiation is not only concentrated on the optical path. The Fresnel zone is an area where the electric field strength varies with the effect from obstacles such as blockings, diffractions, and reflections. Technically, the LoS propagation is cut only when the Fresnel zone is totally blocked by obstructions. The ellipsoid equation of Fresnel clearance zone can be expressed as \cite{Liu18_CL, Cui20_ITJ}
\setcounter{equation}{1}
\begin{equation}
\dfrac{{{x}^{2}}}{{{X}^{2}}}+\dfrac{{{y}^{2}}}{{{Y}^{2}}}+\dfrac{{{z}^{2}}}{{{Z}^{2}}}\le 1
\label{2}
\end{equation}
\noindent and the ellipse parameters can be calculated by
\setcounter{equation}{2}
\begin{equation}
\left\{
\begin{array}{l}
X=Z=\dfrac{\sqrt{n\lambda {{d}_{\text{RX}}}}}{2} \vspace{2ex}\\
Y=\sqrt{\dfrac{n\lambda {{d}_{\text{RX}}}}{4}+\dfrac{(d_{\text{RX}})^{2}}{\text{4}}}
\label{3}
\end{array}
\right.
\end{equation}
\noindent where $\lambda $ is the wavelength and $n$ is the order index of Fresnel zone. Since the Fresnel zone is the function of wavelength, the LoS probability is also related with the carrier frequency \cite{Liu18_CL, Cui20_ITJ}.
\par According to the Huygens-Fresnel principle, the field strength within the first-order Fresnel zone is the half of total one. In this paper, the LoS propagation is defined that all buildings from the TX to the RX are below the first-order Fresnel ellipsoid. On this basis, the allowed height of building at any point as shown in Fig. 1 can be obtained as
\setcounter{equation}{3}
\begin{equation}
\begin{array}{l}
{{h}^{'}_{\text{LoS}}}={{h}_{\text{LoS}}}-\Delta h \vspace{2ex}\\
\ \ \ \ \ \ ={{h}_{\text{TX}}}-\dfrac{{{d}_{\text{LoS}}}({{h}_{\text{TX}}}-{{h}_{\text{RX}}})}{{{d}_{\text{RX}}}}-{{R}_{\text{LoS}}}\cos \theta
\label{4}
\end{array}
\end{equation}
\noindent where $\theta$ is the elevation angle, ${{d}_{\text{LoS}}}$ is the distance from the TX to the obstruction, and ${{R}_{\text{LoS}}}$ can be further calculated by
\setcounter{equation}{4}
\begin{equation}
{{R}_{\text{LoS}}}=\left\{
\begin{array}{l}
\dfrac{\sqrt{n\lambda {{d}_{\text{RX}}}}{{d}_{\text{LoS}}}}{{{d}_{\text{RX}}}},{{d}_{\text{LoS}}}\le \dfrac{{{d}_{\text{RX}}}}{2} \vspace{2ex}\\
\dfrac{\sqrt{n\lambda {{d}_{\text{RX}}}}({{d}_{\text{RX}}}-{{d}_{\text{LoS}}})}{{{d}_{\text{RX}}}},{{d}_{\text{LoS}}}>\dfrac{{{d}_{\text{RX}}}}{2} \\
\end{array}
\right.
\label{5}
\end{equation}
\noindent Note that (4)--(5) is general and suitable for the $n$th-order Fresnel zone case, while in the following derivation we only consider $n=1$.
\setcounter{equation}{10}
\begin{figure*}[!t]
\begin{equation}
{{P}_{\text{LoS}}}({{d}_{\text{RX}}},{{h}_{\text{TX}}},{{h}_{\text{RX}}},\psi ,\lambda )=\prod\limits_{i=\text{1}}^{{{N}_{b}}}{\left[ 1-\exp \left( -\dfrac{{{\left[ {{h}_{\text{TX}}}-\dfrac{{{d}_{i}}({{h}_{\text{TX}}}-{{h}_{\text{RX}}})}{{{d}_{\text{RX}}}}-\dfrac{\sqrt{\lambda {{d}_{\text{RX}}}}\min ({{d}_{i}},{{d}_{\text{RX}}}-{{d}_{i}})}{\sqrt{{{d}_{\text{RX}}}^{\text{2}}+{{({{h}_{\text{TX}}}-{{h}_{\text{RX}}})}^{\text{2}}}}} \right]}^{2}}}{2{{\gamma }^{\text{2}}}} \right) \right]}
\label{11}
\end{equation}
\end{figure*}
\section{Geometry-based Stochastic LoS Probability Model}
\subsection{Analytical Stochastic Model}
\par Following the mathematical steps in Section II, we can obtain the LoS probability by the product of probabilities that each building does not obstruct the LoS path. This method relies heavily on the statistical properties of the undergoing scenario. A well-known stochastic classification and description for urban scenarios can be addressed in \cite{Holis08_TAP, ITU-R1410}. It divides the urban scenario into four typical categories, i.e., Suburban, Urban, Dense urban, and High-rise urban. Then, four scenarios are described by three parameters $\psi \in \{\alpha ,\beta ,\gamma \}$, where $\alpha $ is the percent of land area covered by buildings, $\beta $ represents the mean number of buildings, and $\gamma $ denotes the random building height with the probability density function (PDF) as
\setcounter{equation}{5}
\begin{equation}
P(h)=\frac{h}{{{\gamma }^{\text{2}}}}\exp \left( -\frac{{{h}^{2}}}{2{{\gamma }^{2}}} \right).
\label{6}
\end{equation}
\noindent By using the analytical method in \cite{ITU-R1410}, the authors in \cite{Hourani14_WCL} proved that if the straight line is not blocked by any buildings between the TX and the RX, the LoS probability can be expressed as (7).
\noindent It should be mentioned that this result does not consider the factor of building width for simplicity. Each building is approximated by a two-dimensional plane without thickness. Moreover, the Fresnel clearance zone is not considered in this model. These two factors may lead to incorrect prediction result in real applications. To overcome this shortcoming, we take all the factors, i.e., building height distribution, building width, building space, carrier frequency, and transceivers' heights into account and derive a general closed-form expression of LoS probability in the following.
\par The radio propagation has a much wider zone than the straight line. Considering the effect of building width and first-order Fresnel zone, the probability of building shorter than ${{{h}'}_{\text{LoS}}}$  can be rewritten as
\setcounter{equation}{7}
\begin{equation}
\begin{array}{l}
{{P}_{i}}=P({{h}_{i}}<{{h}^{'}_{\text{LoS}}})=\int_{0}^{{h}^{'}_{\text{LoS}}}P(h)\text{d}h \vspace{2ex}\\
\ \ \ \ =1-\exp \left[ -\dfrac{({{h}^{'}_{\text{LoS}}})^{2}}{2{{\gamma }^{2}}} \right]. \\
\label{8}
\end{array}
\end{equation}
\noindent Then, the LoS probability is defined as the product of all ${{P}_{i}}$ with $i$ denoting the index of the possible existing building along the propagation path. Since all buildings and streets are equally distributed \cite{ITU-R1410}, the average number of buildings along the propagation path can be calculated by
\setcounter{equation}{8}
\begin{equation}
\begin{array}{l}
{{N}_{b}}=\text{floor}\left({{d}_{\text{RX}}}\sqrt{\alpha \beta }/1000\right).
\label{9}
\end{array}
\end{equation}
\noindent where floor($x$) represents the downward rounding function. Furthermore, the average building width can be obtained from the measurement data or scenario-dependent parameters as
\setcounter{equation}{9}
\begin{equation}
\begin{array}{l}
W=1000\sqrt{\alpha /\beta}.
\label{10}
\end{array}
\end{equation}
\noindent Finally, by substituting (8)--(10) into (1), the closed-form expression of LoS probability in this paper yields as (11), where ${{d}_{i}}$ is the distance from the TX to the $i$th building as
\setcounter{equation}{11}
\begin{equation}
{{d}_{i}}=\dfrac{(i-0.5){{d}_{\text{RX}}}}{\text{floor(}{{d}_{\text{RX}}}\sqrt{\alpha \beta }/1000)}+\dfrac{W}{2}
.\label{12}
\end{equation}
\noindent As we can see from (11) that the general model is determined by the transceivers' positions and environment-dependent parameters. It includes the factors of building width and first-order Fresnel zone and can be applicable for any heights of TX and RX. Moreover, by setting $W=0$ and ${{R}_{\text{LoS}}}=0$, our model reduces to the traditional result of (7).
\par To observe the effect of transceivers' height on the LoS probability, we take the modern city with high buildings as an example and set the environmental parameter as $\psi =(0.5,300,50)$. Moreover, the rest simulation parameters are listed as follows, ${{h}_{\text{RX}}} = 1.5\text{ m}$, $f = 6\text{ GHz}$ and ${{h}_{\text{TX}}} = 30,\ 300,\ 800 \text{ and } 1500 \text{ m}$. The LoS probabilities with different heights by applying our proposed model are shown in Fig.~2. It clearly shows that the LoS probability increases with the height of TX, which is apparently reasonable from the intuition. For the purpose of mobile communication system design, the LoS path is vital to guarantee the performance threshold. As an example, if the required LoS probability is set as 0.6, we can obtain the average maximum communication distance (MCD) by the aerial base station. As we can see from Fig.~2, the MCDs are 70.2~m, 157.6~m, 305.9~m, and 515.4~m for different heights, respectively. The similar result can be found in \cite{Cui20_ITJ}, which is about 70~m at the TX height of 25~m with the threshold of 0.6. These analytical results are helpful for the coverage radius and layout optimization in UAV-aided communications systems.
\begin{figure}[!t]
	\centering
	\includegraphics[width=85mm]{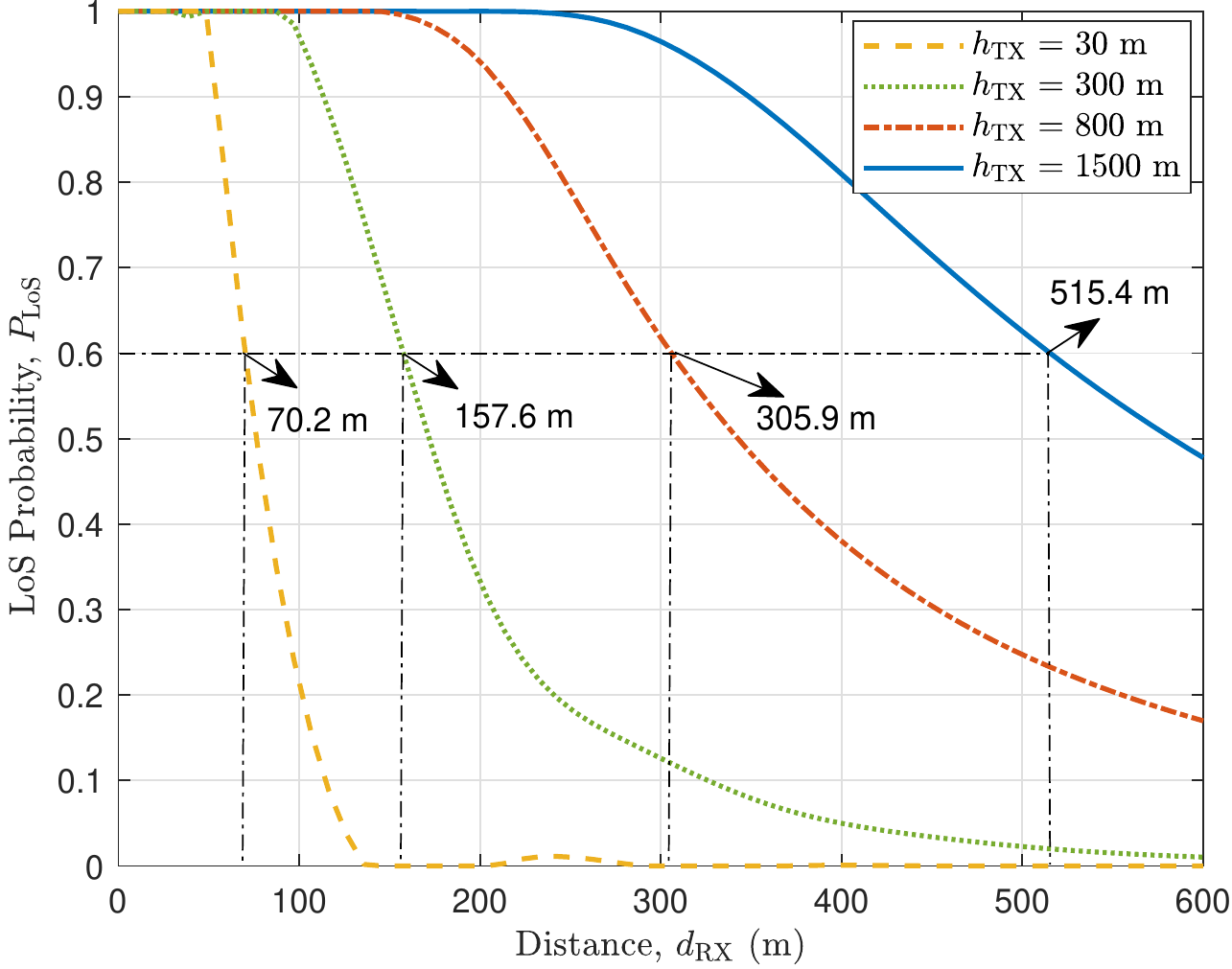}
	\caption{MCD at different communication heights.}
    \label{fig:2}
\end{figure}
\par To analyze the effects of building width and Fresnel zone on the LoS probability, Fig.~3 compares the probabilities with different widths and frequencies. In the simulation, we set the building width as 0~m, 20~m, and 40~m, respectively, and the frequency as 1.2~GHz, 6~GHz, 28~GHz, and infinite, respectively. The rest simulation parameters are listed as follow, ${{h}_{\text{TX}}}=70\text{ m}$, ${{h}_{\text{RX}}}=1.5\text{ m}$, and $\gamma = 15$. As it is shown in Fig.~3 that the LoS probability declines when the building width increases. The main reason is that the wider building would increase the possibility of the blocking to the LoS path. Meanwhile, the lower frequency leads to smaller LoS probability, because the corresponding radius of Fresnel zone is larger and easier to be obstructed. We also find that the impact of frequency tends to be unapparent when the frequency is up to 6~GHz. It is reasonable because the radius difference of first-order Fresnel zone is very small at the high frequency.
\begin{figure}[!b]
	\centering
	\includegraphics[width=85mm]{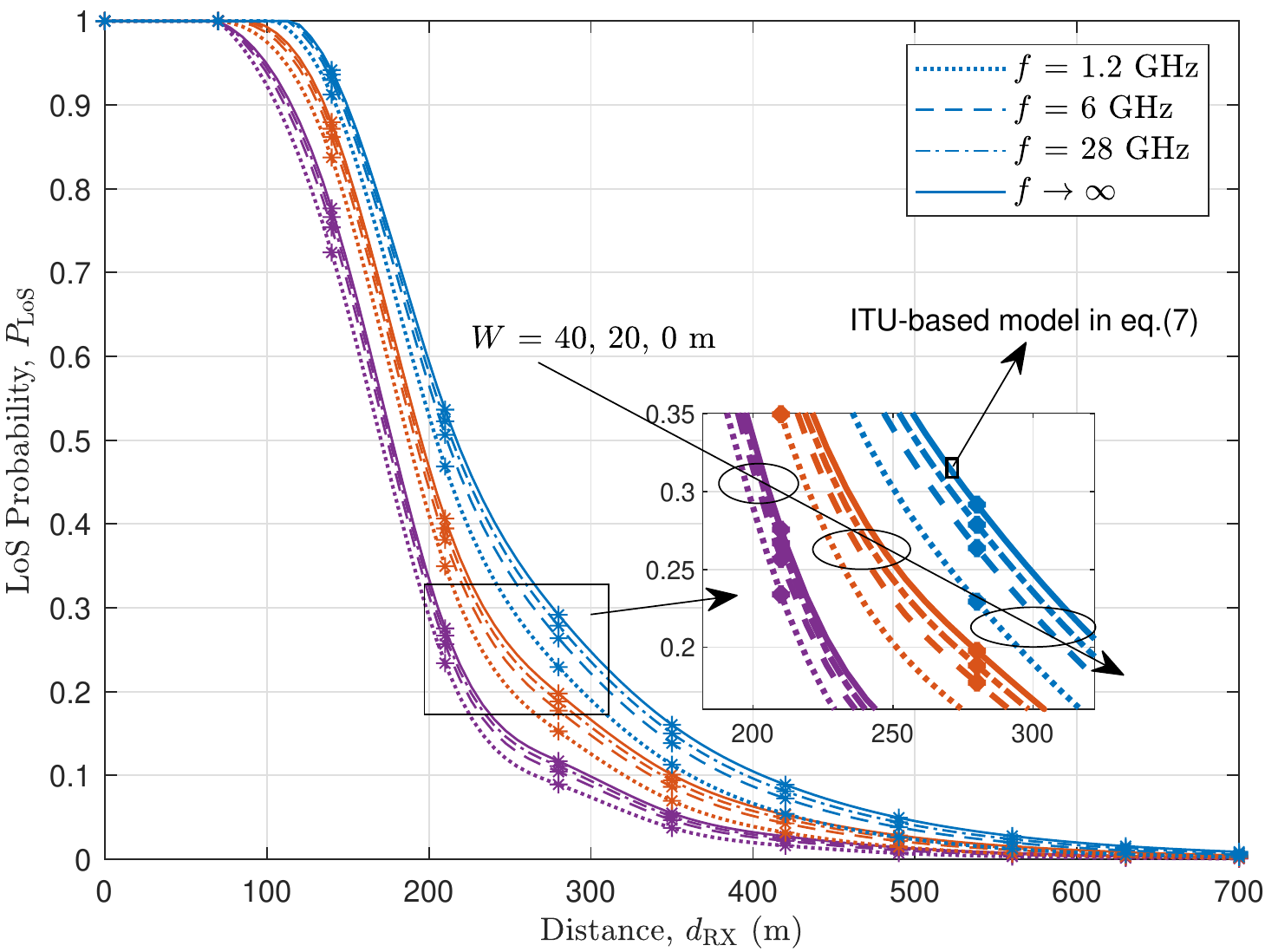}
	\caption{Effects of building width and frequency on the LoS probability.}
    \label{fig:3}
\end{figure}
\setcounter{equation}{12}
\begin{figure*}[!t]
\begin{equation}
\begin{array}{l}
{{P}_{\text{LoS}}}({{d}_{\text{RX}}})=\min \left( \dfrac{{{D}_{1}}}{{{d}_{\text{RX}}}},1 \right)\cdot \left[ 1-\exp \left( -\dfrac{{{d}_{\text{RX}}}}{{{D}_{2}}} \right) \right] +\exp \left( -\dfrac{{{d}_{\text{RX}}}}{{{D}_{2}}} \right)
\label{13}
\end{array}
\end{equation}
\end{figure*}
\setcounter{equation}{13}
\begin{figure*}[!t]
\begin{equation}
\begin{array}{l}
D_{1/2}^{(n)}\left( \Delta h,\mathbf{w},\mathbf{b},\mathbf{\varepsilon } \right)=\varepsilon _{1}^{\left( 2 \right)}\left[ \sum\limits_{j=1}^{J}{w_{j1}^{\left( 2 \right)}\varepsilon _{j}^{\left( 1 \right)}\left( w_{1j}^{\left( 1 \right)}\Delta {{h}^{(n)}}+b_{j}^{\left( 1 \right)} \right)+b_{1}^{\left( 2 \right)}} \right]
\label{14}
\end{array}
\end{equation}
\end{figure*}
\begin{figure*}[!b]
	\centering
	\includegraphics[width=160mm]{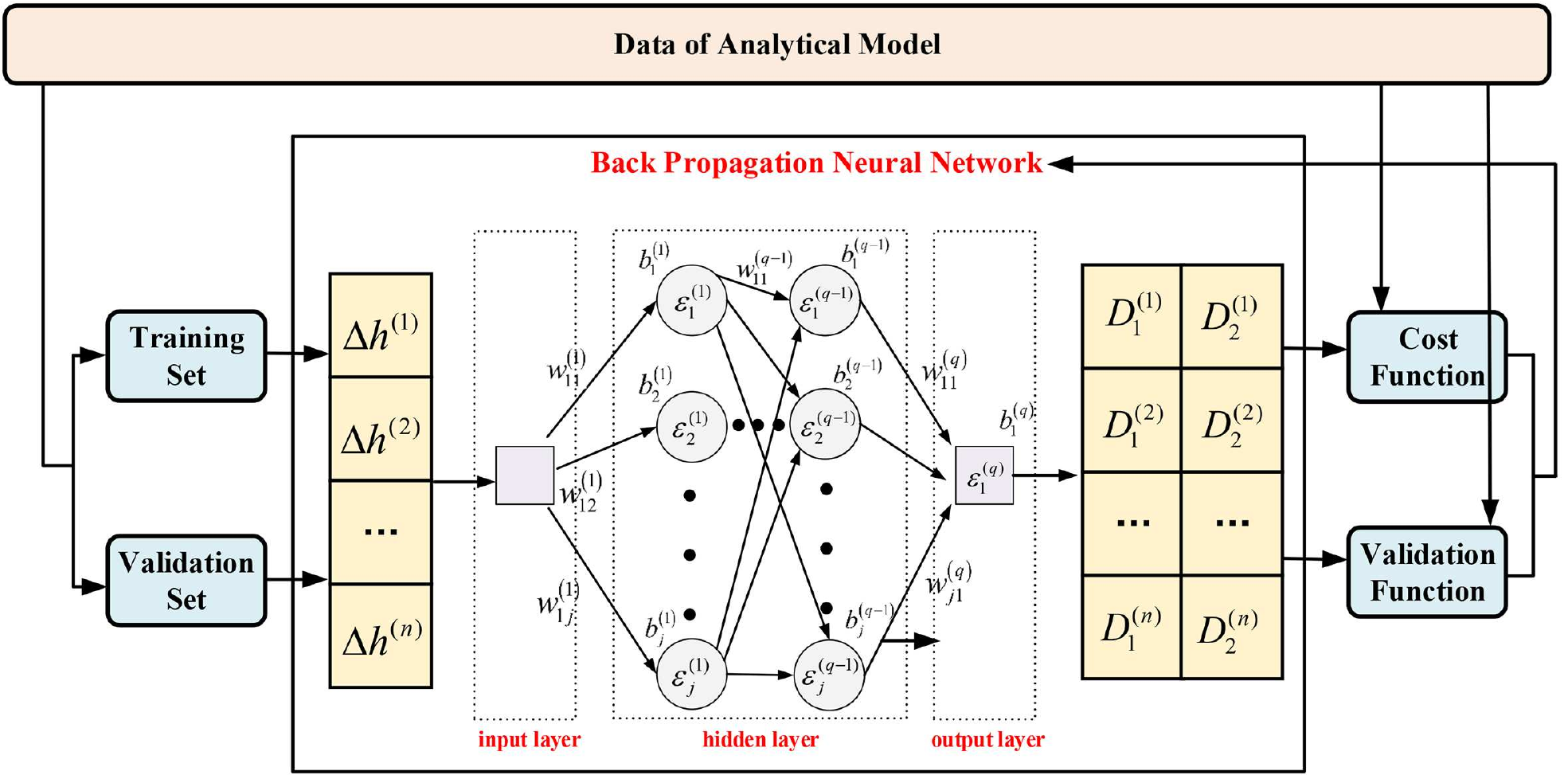}
	\caption{BPNN-based framework for parameter generation.}
    \label{fig:4}
\end{figure*}
\subsection{Approximate Model with ML-based Parameters}
\par The analytical model of LoS probability is general and accurate, but it requires detailed environment information and is time-consuming. Especially, for the purpose of deriving a closed-form solution of system performance and channel model, we prefer an approximate but closed-form expression to calculate the LoS probability. In the literature, an alternative modeling approach based on real measurement data, namely empirical method, was addressed in \cite{ITU-R2135, 3GPP, 5GCM}. Different from the geometric method, the empirical method is aimed to obtain a simple and easy-to-use prediction model for specific scenes. Note that the existing empirical models of LoS probability focused on the land mobile communications. The transceivers usually represent the base station and mobile station with the heights of 25~m and 2~m, respectively. Since the measured results are significantly different for different heights of transceivers, these empirical models cannot be directly applied to A2G communications. Moreover, massive field measurements for A2G scenarios are high cost and difficult, so the analytical results of proposed model are used instead of it in this paper.
\par The measurement results have shown that the LoS probability is much more related with the exponent of distance than frequency band. The 3GPP-based LoS probability model can be expressed as (13) in \cite{ITU-R2135, 3GPP, 5GCM},where ${{d}_{\text{RX}}}$ is the horizontal distance from the TX to the RX, ${{D}_{1}},{{D}_{2}}$ are the parameters fitted by the measurement results. For the land mobile scenarios, ${{D}_{1}}=18\text{ m}$, ${{D}_{2}}=63\text{ m}$ and ${{D}_{1}}=20\text{ m}$, ${{D}_{2}}=66\text{ m}$ are recommended in \cite{ITU-R2135, 3GPP} and \cite{5GCM}, respectively. However, the applicable height of both models is limited within a few tens of meters, which means low-altitude scenarios. In order to overcome this shortcoming, we develop this model to fit the A2G scenario by taking the height factor into account in the following. In (13), ${{D}_{1}}$ is the breakpoint distance where LoS probability is no longer equal to 1, and ${{D}_{\text{2}}}$ is a decay parameter. Note that the LoS probability is described by the distance and the transceiver height difference $\Delta h={{h}_{\text{TX}}}-{{h}_{\text{RX}}}$.
\par As we know, ML technologies can accurately find the inner relationship from massive data. In order to obtain ${{D}_{1}}$ and ${{D}_{\text{2}}}$ as height-dependent parameters, we propose a parameter training and estimation method with a BPNN-based training framework to fit the analytical data. The BPNN-based training framework is given in Fig.~4. The data sets $\{ D_{1}^{(n)},\Delta {{h}^{(n)}}\}$ and $\{ D_{2}^{(n)},\Delta {{h}^{(n)}}\}$ are respectively divided into two parts by the proportion of 7:3 randomly, and 70\% of data is used to train while the other is used to validate. Noted that these two sets have no intersection. The neural network consists of input layer, hidden layer, and output layer. For example, we take the neural network with one input layer $\Delta {{h}^{(\text{1})}}$, one hidden layer of $J$ neurons and one output layer, the ${{D}_{1}}$ and ${{D}_{2}}$ can be expressed as (14), where $w_{ij}^{\left( q \right)}\in \mathbf{w}$ is the connection weight of two neurons, $b_{j}^{\left( q \right)}$ is the bias of the $j$th neuron in the $q$th layer, $\varepsilon _{j}^{\left( q \right)}$ is the activation function of the $j$th neuron in the $q$th layer. In this paper, the sigmoid function is applied in hidden layers and the output cost function is a vital item. We can obtain the best fitting results by adjusting the network parameters to minimize the cost function. Therefore, we design a cost function based on the mean square error (MSE) as
\setcounter{equation}{14}
\begin{equation}
\begin{array}{l}
{{C}_{F}}\left( \mathbf{w},\mathbf{b},\mathbf{\varepsilon } \right)=\dfrac{1}{N}\sum\limits_{n=1}^{N}{{{\left[ D_{1/2}^{(n)}\left( \Delta {{h}^{(n)}} \right)-D_{1/2}^{(n)} \right]}^{2}}} \vspace{2ex}\\
\ \ \ \ \ \ \ \ \ \ \ \ \ \ \ \ +\dfrac{1}{2}\eta \sum\limits_{q=1}^{Q}{{{\left\| \mathbf{w}_{{}}^{\left( q \right)} \right\|}^{2}}}
\label{15}
\end{array}
\end{equation}
\noindent where $\eta $ is the regularization parameter. After training the network, we use the verification function root mean square error (RMSE) to evaluate the network performance after inputting the validation data set.
\begin{figure*}[!t]
\centering
\subfigure[Analytical result of LoS probability]{
\includegraphics[width=80mm]{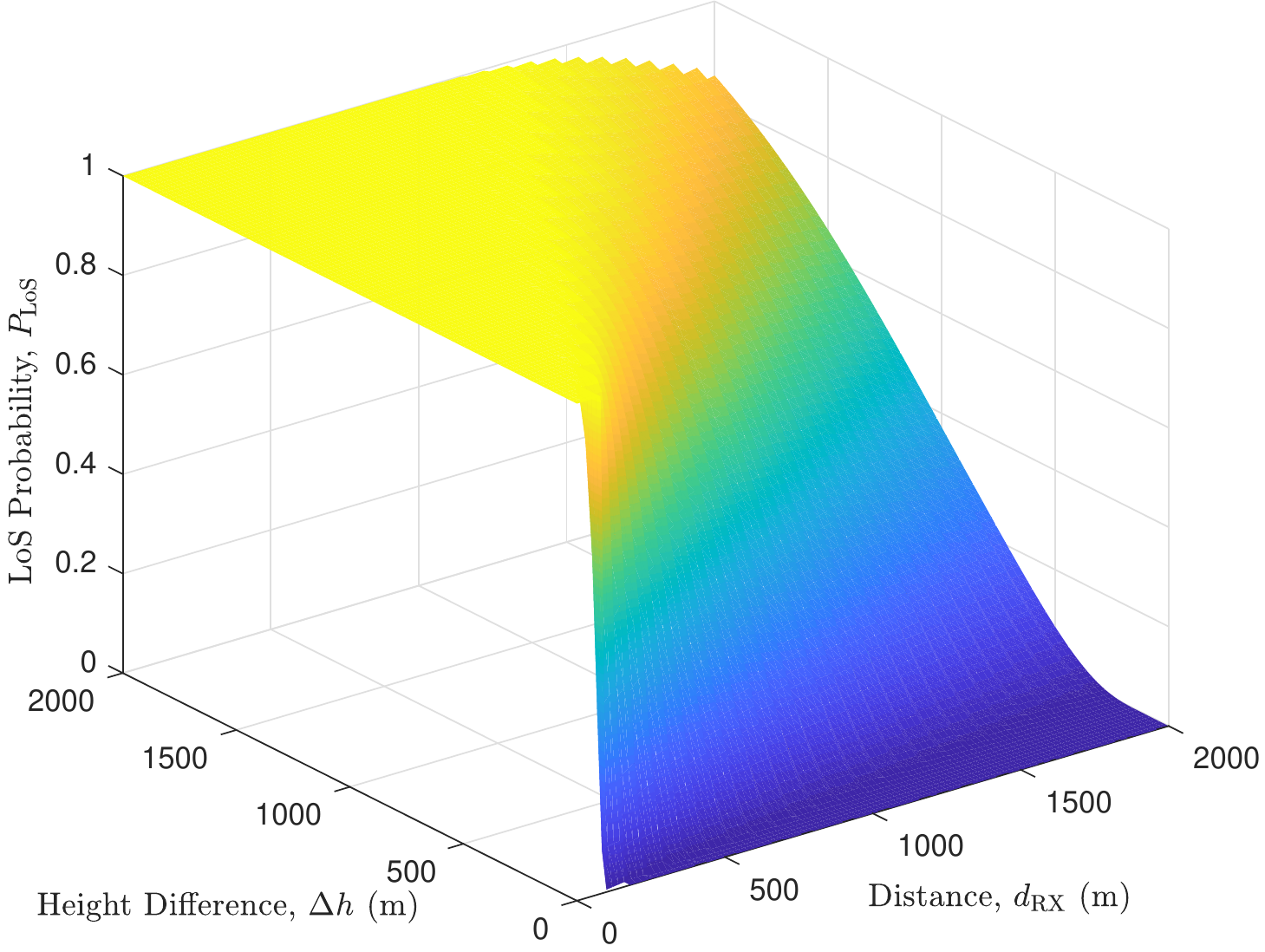}
}
\subfigure[Absolute error of approximate model]{
\includegraphics[width=80mm]{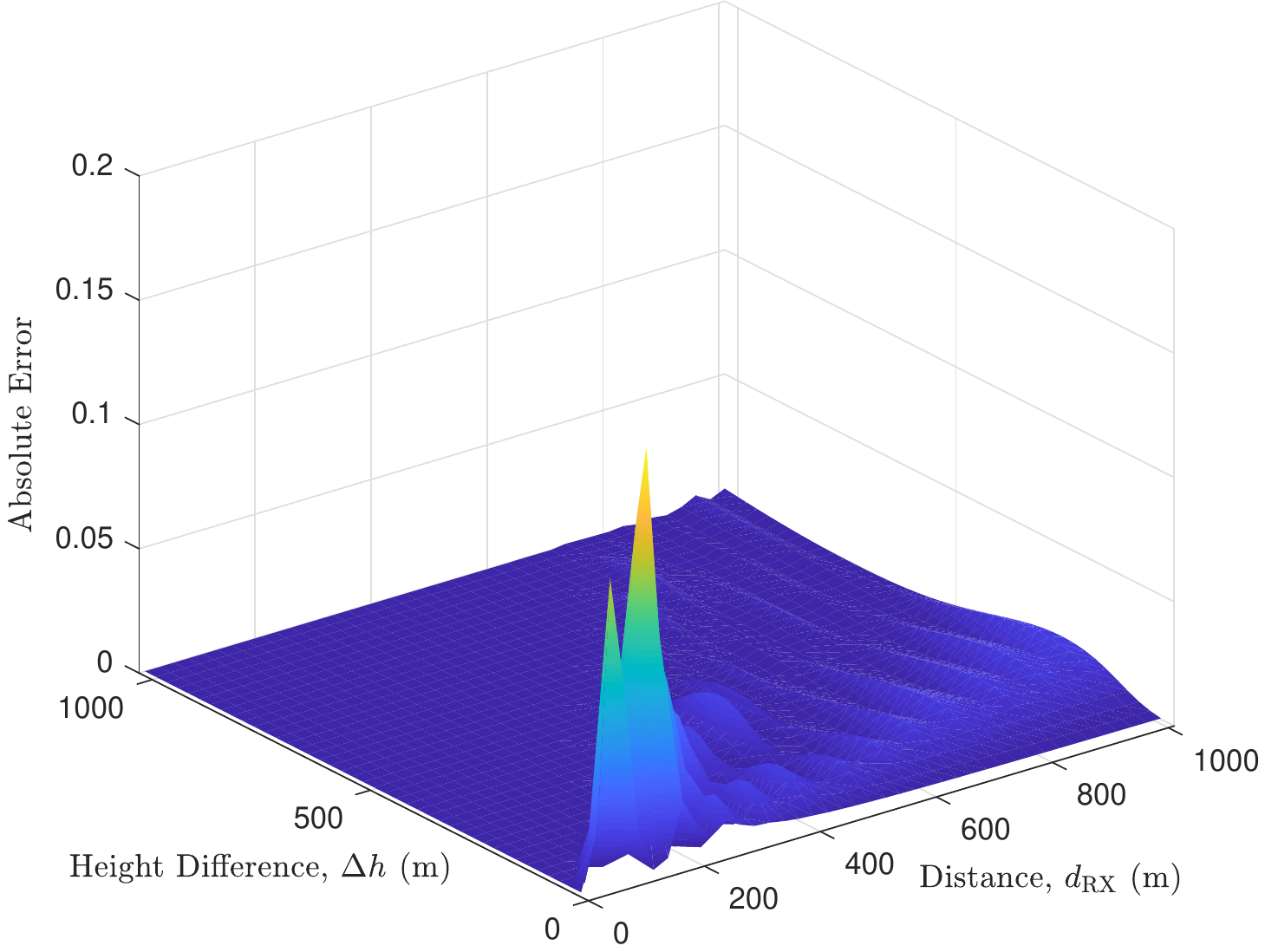}
}
\caption{LoS probability under the urban scenario.}
\end{figure*}
\par Four typical urban environments, i.e., Suburban, Urban, Dense urban, and High-rise urban, are studied in this paper and their environment parameters $\psi \in \{\alpha ,\beta ,\gamma \}$ are (0.1, 750, 8), (0.3, 500, 15), (0.5, 300, 20), and (0.5, 300, 50), respectively. By fitting the simulation result of analytical model with different environment parameters, the trained parameters of BPNN network for upgraded parameters ${{D}_{1}}$, ${{D}_{2}}$ including the height factor are given in Table I.
\begin{table*}[!t]
\renewcommand\arraystretch{1.5}
\centering
\caption{The trained parameters of neurons for different urban scenarios.}
\centering
\begin{tabular}{|p{1cm}<{\centering}|p{1.2cm}<{\centering}|p{1.6cm}<{\centering}|p{1cm}<{\centering}|p{1.5cm}<{\centering}|p{1.5cm}<{\centering}|p{1.8cm}<{\centering}|p{2.2cm}<{\centering}|}
\hline
\multicolumn{4}{|p{4.8cm}<{\centering}|}{Scenario}&{Suburban}&{Urban}&{Dense urban}&{High-rise urban}\\
\hline
\multirow{10}{1cm}{$D_1$} & \multirow{5}{1.2cm}{weight} & \multirow{4}{1.8cm}{hidden lay} & $w_{11}^{\left( 1 \right)}$ & 16.2579 & -1.6587 & 2.4455 & 1.2291 \\ \cline{4-8}
 &  &  & $w_{12}^{\left( 1 \right)}$ & -5.5254 & 5.1759 & -3.5892 & -0.3727 \\ \cline{4-8}
 &  &  & $w_{13}^{\left( 1 \right)}$ & 15.4283 & 9.1645 & 2.5314 & 3.0045 \\ \cline{4-8}
 &  &  & $w_{14}^{\left( 1 \right)}$ & 9.6738 & 4.9191 & 5.2872 & -0.7202 \\ \cline{3-8}
 &  & {output lay} & $w_{11}^{\left( 2 \right)}$ & 5.1456 & 3.4246 & 3.9771 & 2.6658 \\ \cline{2-8}
 &  \multirow{5}{1.2cm}{bias} & \multirow{4}{1.8cm}{hidden lay} & $b_{1}^{\left( 1 \right)}$ & -3.0018 & -1.2296 & -2.7575 & -1.7200 \\ \cline{4-8}
 &  &  & $b_{2}^{\left( 1 \right)}$ & -1.3995 & -3.2076 & -0.8322 & -1.1132 \\ \cline{4-8}
 &  &  & $b_{3}^{\left( 1 \right)}$ & -0.8644 & -1.7848 & -2.7720 & -2.1148 \\ \cline{4-8}
 &  &  & $b_{4}^{\left( 1 \right)}$ & 1.0262 & -3.1057 & -1.3142 & -1.0177 \\ \cline{3-8}
 &  & {output lay} & $b_{1}^{\left( 2 \right)}$ & -6.6230 & -1.4798 & -2.3955 & -1.9291 \\
\hline
\multirow{10}{1cm}{$D_2$} & \multirow{5}{1.2cm}{weight} & \multirow{4}{1.8cm}{hidden lay} & $w_{11}^{\left( 1 \right)}$ & 6.5142 & -13.0707 & 4.6853 & -2.3706 \\ \cline{4-8}
 &  &  & $w_{12}^{\left( 1 \right)}$ & -10.6197 & 8.4525 & 0.3355 & 6.1472 \\ \cline{4-8}
 &  &  & $w_{13}^{\left( 1 \right)}$ & -3.9213 & -1.3332 & 5.7374 & 1.0547 \\ \cline{4-8}
 &  &  & $w_{14}^{\left( 1 \right)}$ & -0.6352 & 7.2757 & -7.3002 & 3.8038 \\ \cline{3-8}
 &  & {output lay} & $w_{11}^{\left( 2 \right)}$ & 3.1422 & 4.1644 & 3.1653 & 1.3593 \\ \cline{2-8}
 &  \multirow{5}{1.2cm}{bias} & \multirow{4}{1.8cm}{hidden lay} & $b_{1}^{\left( 1 \right)}$ & -0.2573 & 2.8829 & -0.1744 & 1.1160 \\ \cline{4-8}
 &  &  & $b_{2}^{\left( 1 \right)}$ & 0.7117 & -0.5461 & -0.8997 & -0.8216 \\ \cline{4-8}
 &  &  & $b_{3}^{\left( 1 \right)}$ & -2.7229 & -1.9083 & 0.3100 & 1.6107 \\ \cline{4-8}
 &  &  & $b_{4}^{\left( 1 \right)}$ & -1.4552 & 0.1436 & 2.9326 & -0.3733 \\ \cline{3-8}
 &  & {output lay} & $b_{1}^{\left( 2 \right)}$ & -2.8366 & -7.8225 & -6.7432 & -2.6654 \\
 \hline
\end{tabular}
\label{table I}
\end{table*}
\par The absolute error of approximate model under the urban scenario with $\psi =(0.3,500,15)$ is evaluated in the following. The range of both ${{d}_{\text{RX}}}$ and $\Delta h$ are observed from 0~m -- 1000~m. As we can see in Fig.~5(a) that the LoS probability declines with the increasing distance, and the larger height difference would increase the existing probability of LoS path. Fig.~5(b) shows that the absolute error between the analytical model and approximate model is from 0.01 -- 0.09. Moreover, we can obtain that the MSE is less than 0.03. The good consistency between analytical and approximate results shows the applicability of the simplified parametric model. Note that it could be quite different in various real environments since the building distribution is too complicated to be described by three parameters. Besides, the additional possible block from trees, lamp-posts, and mobile obstacles are neglected in the analytical model. Thus, it is not trivial for the prediction error due to the approximation. In fact, this approximate model with lower complexity and closed-form expression may have more wide application in the future channel modeling and communication performance evaluation.
\section{Simulation Results and Validations}
\subsection{RT-based Monte-Carlo Simulations}
\par Measurement campaigns for A2G channels are difficult and high cost. Moreover, in order to check the validity of stochastic model, a large number of measurements are needed. The RT technique has been widely applied to channel modeling and validation to alleviate the burden of measurement campaigns \cite{Taygur20_TAP}. In this section, the RT-based Monte-Carlo simulations are conducted to validate the proposed LoS probability model. For RT simulation, the electromagnetic wave is considered to be a bunch of rays, so a geometric solution can be obtained based on the uniform theory of diffraction and geometric optics. After tracking all rays, a lot of propagation parameters including the existing of LoS path can be obtained.
\par The RT-based Monte-Carlo simulation mainly includes two steps, i.e., scenario reconstruction and probability computation. Firstly, by using the parameters of building width, building height, and building space according to the environmental clarification, we can reconstruct the virtual scenario with the same statistical characteristic as shown in Fig.~6. Secondly, the probability computation can be conducted by three steps, i.e., decomposition of ray source, tracking rays, and intersection operation. If there is no obstacle in the first-order Fresnel zone, the arrival ray can be counted as the LoS path. In general, the existence of LoS path determined by the intersection of rays and the first Fresnel zone, which can be transformed to the solution of triangular surface intersection as shown in Fig.~7. The detailed solution of intersection operation can be found in the Appendix.
\begin{figure}[!b]
	\centering
	\includegraphics[width=85mm]{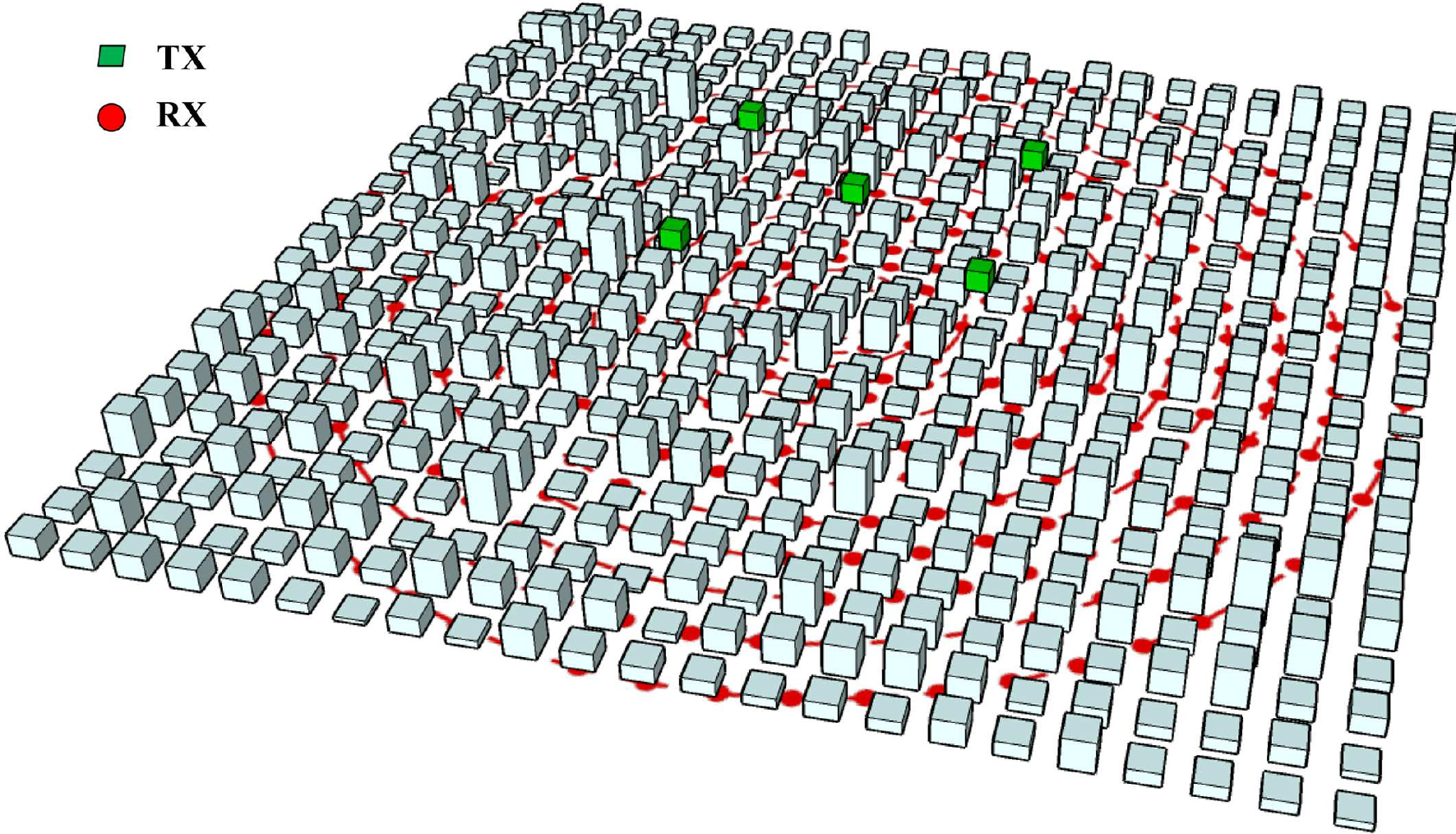}
	\caption{Reconstructed urban scenario.}
    \label{fig:6}
\end{figure}
\begin{figure}[!t]
	\centering
	\includegraphics[width=40mm]{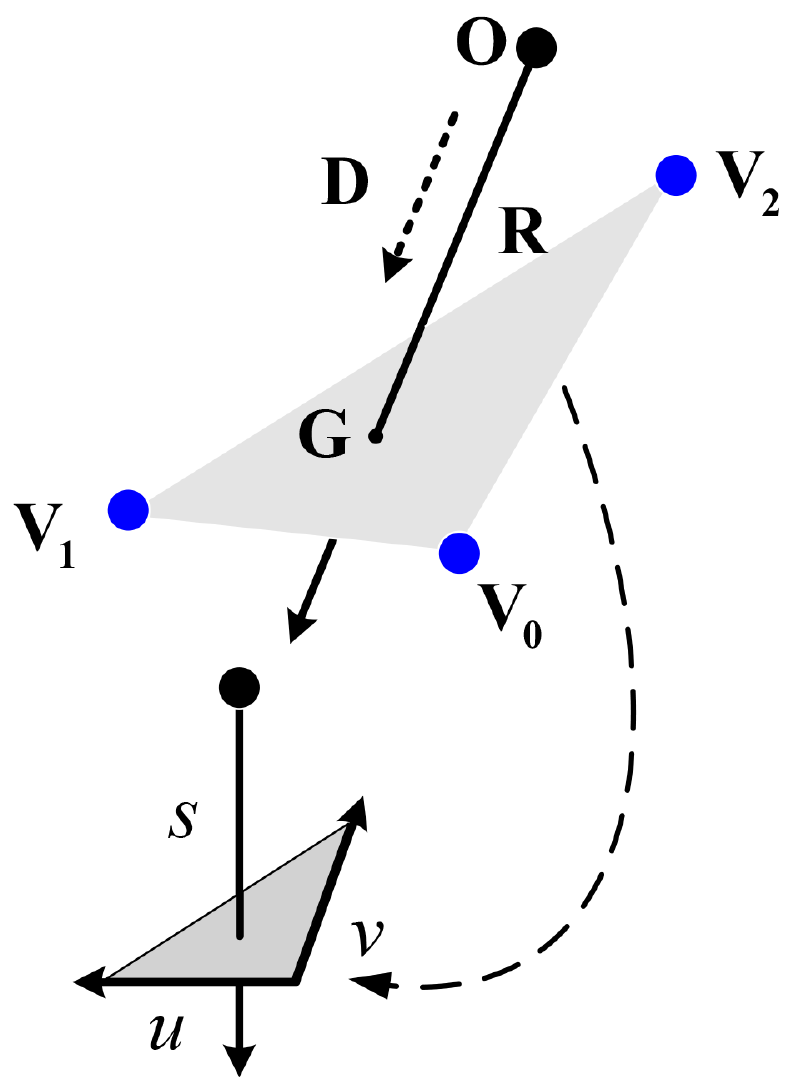}
	\caption{Intersection operation.}
    \label{fig:7}
\end{figure}
\subsection{Simulation Results and Analysis}
\par In this part, we provide several numerical simulation results and comparisons to demonstrate the accuracy of our proposed model for different heights and distances. Following the above steps, we reconstruct the virtual city scenarios including standard Urban, Dense urban, and High-rise urban, according to the environmental parameters as (0.3,500,15), (0.5,300,20) and (0.5,300,50). Building heights are randomly generated based on the distribution, while the width and space are kept constant for each generated scenario. The mean value of building width in these three scenarios are 24.5~m, 40.8~m, and 40.8~m respectively. Moreover, five TXs at the height of 500~m are located in the center, and the RXs are uniformly distributed on several concentric circles with the height of 2~m. The simulated LoS probability is obtained by dividing the number of LoS pairs by the total number of transceiver pairs with the same height and distance. The detailed parameters can be found in Table II.
\par For some applications, LoS probability is described as a function of elevation angle at high latitudes for more intuitive observation \cite{Holis08_TAP, Hourani14_WCL}. We set the elevation angle as $\theta =\arctan ({h}'/d)$ and transform the proposed model with respect to $\theta $ as shown in Fig.~8. The good consistency between the proposed model and simulated data shows that our prediction model has satisfied applicability under different urban scenarios. It also demonstrates that the denser and higher buildings layout would improve the block possibility. Moreover, when the LoS probability threshold is set as 0.6, the minimum values of elevation angle under standard Urban, Dense urban, and High-rise urban are $\text{32}\text{.}{{\text{5}}^{\circ }}$, ${{50.6}^{\circ }}$ and ${{72.6}^{\circ }}$, which are also similar with the results in \cite{Holis08_TAP} as ${{25}^{\circ }}$,${{50}^{\circ }}$ and ${{71}^{\circ }}$ for the high altitude case.
\begin{figure}[!b]
	\centering
	\includegraphics[width=85mm]{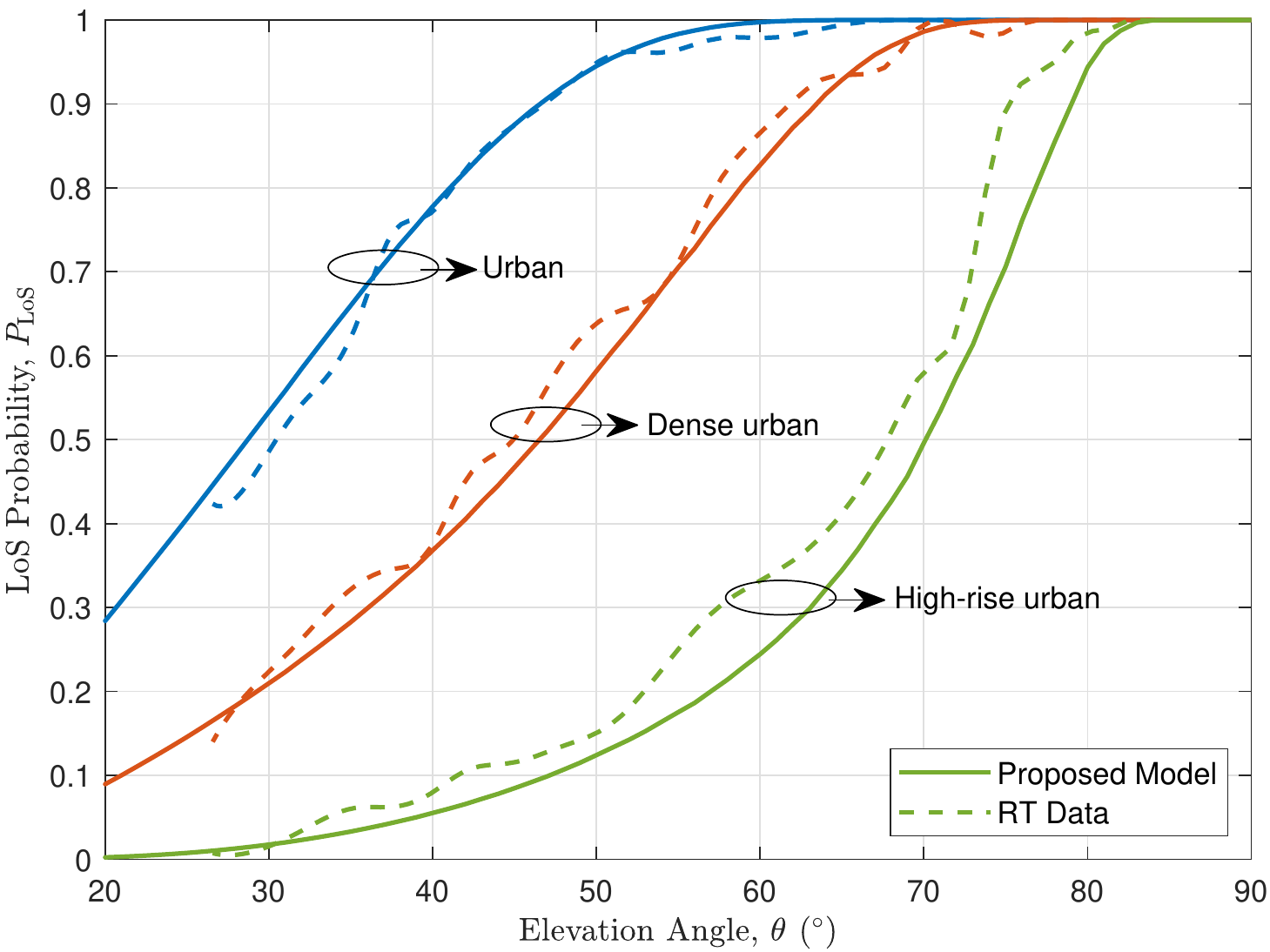}
	\caption{LoS probability with respect to the elevation angle for urban scenarios.}
    \label{fig:8}
\end{figure}
\begin{table*}[!t]
\renewcommand{\arraystretch}{1.3}
\caption{RT Simulation parameters}
\label{table_2}
\centering
\begin{tabular}{p{4.4cm}p{5.6cm}}
\hline
Parameter & Value\\
\hline
Scenario & Urban, Dense urban and High-rise urban\\
Building location distribution & Uniform distribution\\
Building width & 24.5 m, 40.8 m and 40.8 m\\
Building height distribution & Rayleigh distribution\\
Building height & 15 m, 20 m and 50 m\\
Carrier frequency & 28 GHz\\
Antenna type & Omnidirectional\\
TX height & 500 m\\
TX number & 5\\
RX height & 2 m\\
RX number & 5256\\
\hline
\end{tabular}
\end{table*}
\par To further evaluate the prediction performance of proposed model, we compare the predicted results with the ones of other existing models, e.g., 3GPP, NYU, the model in \cite{Hourani20_WCL}, as well as the RT-based Monte-Carlo method. The RX height is set as 2~m, while the TX heights are set as 30~m, 120~m, and 500~m, representing the low altitude (traditional base station), middle altitude (typical UAVs), and high altitude (other HAPs), respectively.
\begin{figure}[!t]
\centering
\subfigure[${{h}_{\text{TX}}}=30\text{ m}$]{
\includegraphics[width=80mm]{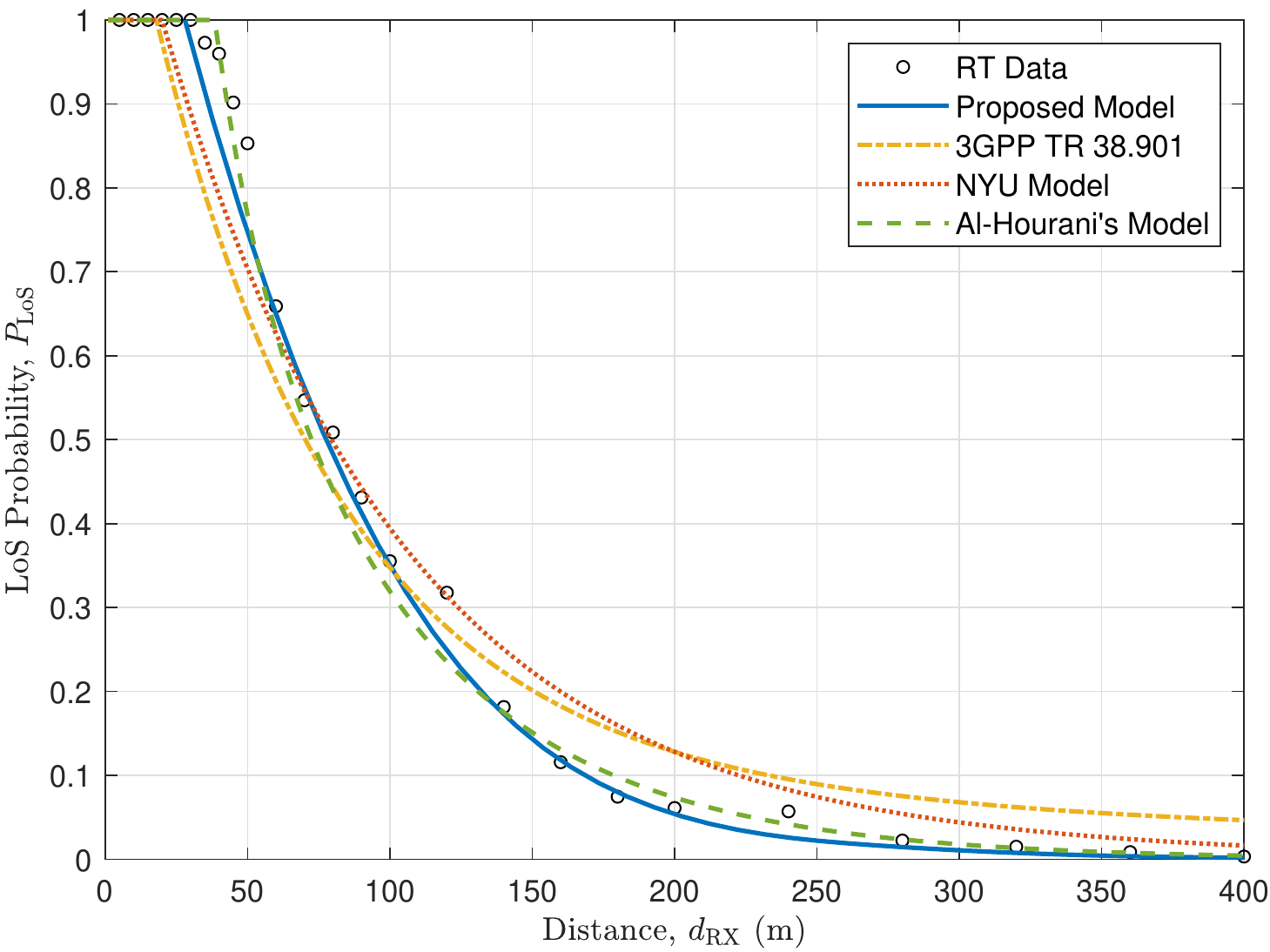}
}
\subfigure[${{h}_{\text{TX}}}=120\text{ m}$]{
\includegraphics[width=80mm]{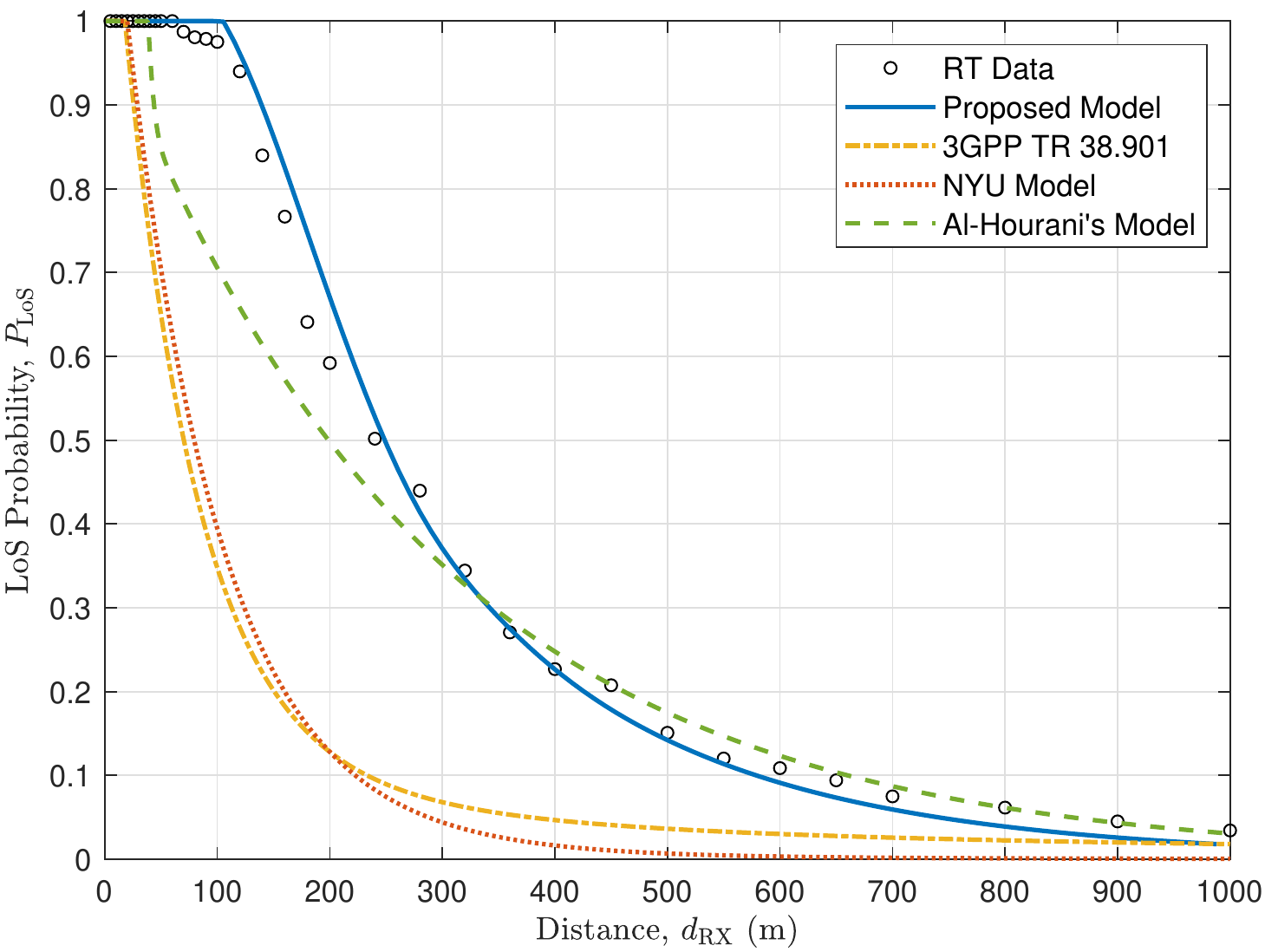}
}
\subfigure[${{h}_{\text{TX}}}=500\text{ m}$]{
\includegraphics[width=80mm]{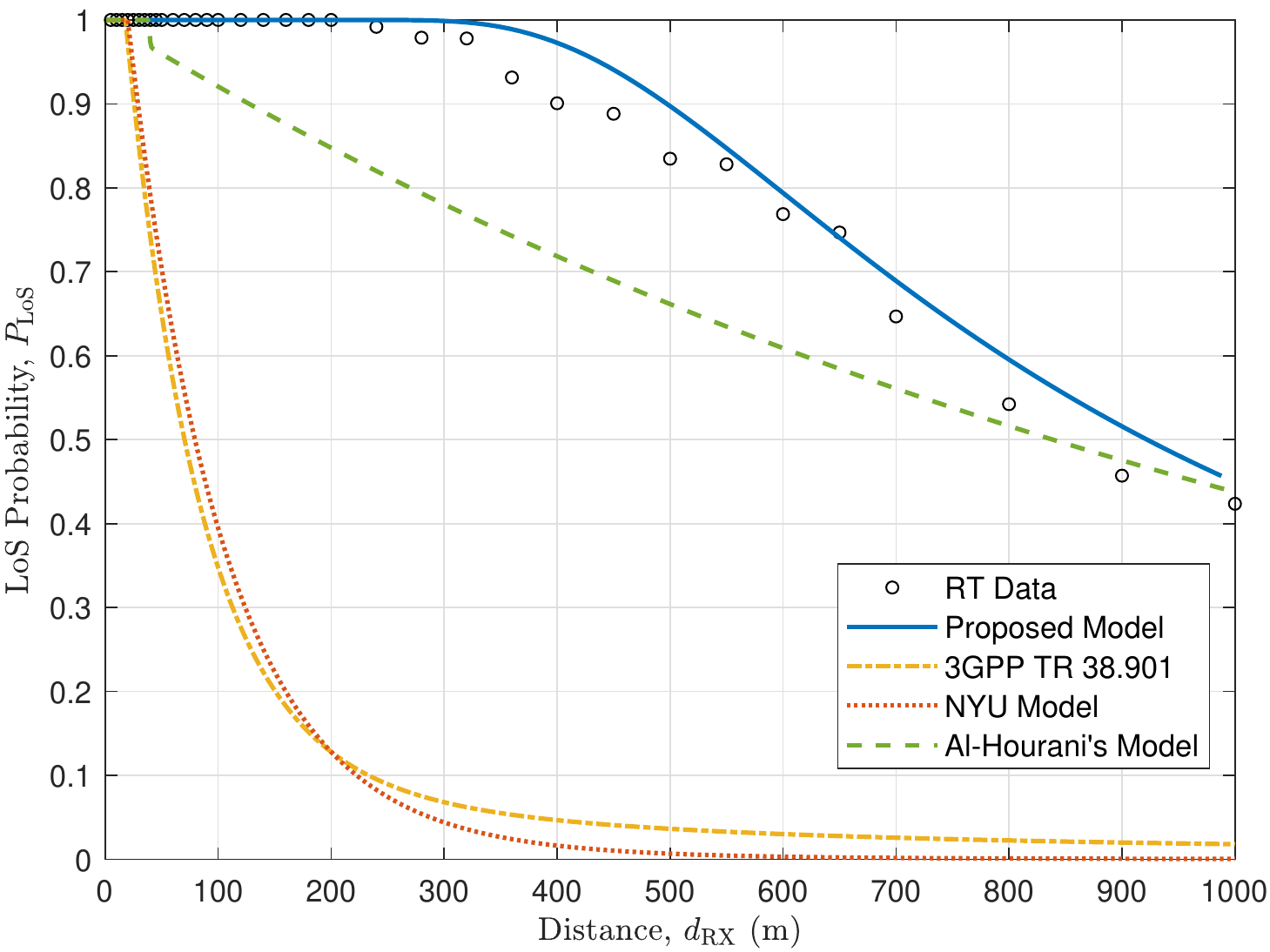}
}
\caption{Comparison of proposed models with other models and RT data.}
\end{figure}
\par Without loss of generality, the Urban scenario is chosen and the comparison results are shown in Fig.~9. For the case of low altitude as shown in Fig.~9(a), the good consistency of all models and Monte-Carlo method shows that our model can achieve effective prediction. Moreover, the breakpoint distance where the LoS probability is no longer equal to 1 is found to concentrate in the range of 20~m--40~m. The prediction results of the Model in \cite{Hourani20_WCL} and our model are a little bigger than those of the other models. It makes sense since the 3GPP and NYU model are based on the measurement data which includes the effect of other omitted scatters, e.g., pedestrian, cars, and trees. It's worth to mentioning that the 3GPP and NYU model are only suitable for low altitude case (tens of meters) due to the measurement results. When the TX height comes to 120~m, their prediction results deviate from the simulated data as shown in Fig.~ 9(b). Especially, the breakpoint distances of them are almost unchanged which is obviously unreasonable. Theoretically, the breakpoint distance should show an increasing trend as the height rises. However, the breakpoint distance of the model in \cite{Hourani20_WCL} shows very few changes even for the high altitude, which seems also not be acceptable. As depicted in Fig.~9(c), the error between the RT data and other models become obvious when the TX height reaches 500~m. In all cases, the prediction results of our proposed model agree well with the those of the RT-based simulation data, and also show the increasing trend of breakpoint distance as the height rises, i.e., 30~m, 100~m , and 300~m at three different altitudes.

\section{Conclusions}
\par In this paper, we have derived a general expression for LoS probability of A2G channels by using the statistical characteristic and geometric information of urban scenarios. In addition, a simplified parametric model has been proposed to speed up the prediction and assist in the closed-form solution for channel modeling and system performance. Based on the analytical stochastic model, a BPNN-based framework has been developed and trained for parameter generation. The simulation results have shown that the proposed model demonstrates satisfied versatility at low, middle, and high altitudes and has a good agreement with RT-based Monte-Carlo simulated data. The proposed theoretical and simplified models have a wide range of applications such as channel modeling, link budget, and system optimization for the A2G communications. Future work will include the channel measurement of LoS probability and incorporating the factor of antenna pattern.
\appendix
\section{}
\par Each obstacle can be reconstructed by several triangle facet, and we denote each target triangle facet with three point $u$, $v$ and $s$ as shown in Fig.~7. The location of arbitrary point $\mathbf{G}$ on the triangle facet can be expressed as \cite{Moller05}
\setcounter{equation}{15}
\begin{equation}
\mathbf{G}(u,v)={{\mathbf{V}}_{\mathbf{0}}}+u({{\mathbf{V}}_{\mathbf{1}}}-{{\mathbf{V}}_{\mathbf{0}}})+v({{\mathbf{V}}_{\mathbf{2}}}-{{\mathbf{V}}_{\mathbf{0}}})
\label{16}
\end{equation}
\noindent where ${{\mathbf{V}}_{\mathbf{0}}}$, ${{\mathbf{V}}_{\mathbf{1}}}$, and ${{\mathbf{V}}_{\mathbf{2}}}$ are the three vertices of the triangle. A ray $\mathbf{R}(s)$ with origin $\mathbf{O}$ and normalized direction $\mathbf{D}$ is defined as
\setcounter{equation}{16}
\begin{equation}
\mathbf{R}(s)=\mathbf{O}+s\mathbf{D}
\label{17}
\end{equation}
\noindent The intersection between the ray and the triangle equals to $\mathbf{R}(t)=\mathbf{G}(u,v)$, which yields
\setcounter{equation}{17}
\begin{equation}
\left[ \begin{matrix}
   -\mathbf{D} & {{\mathbf{V}}_{\mathbf{1}}}-{{\mathbf{V}}_{\mathbf{0}}} & {{\mathbf{V}}_{\mathbf{2}}}-{{\mathbf{V}}_{\mathbf{0}}}  \\
\end{matrix} \right]\left[ \begin{matrix}
   s  \\
   u  \\
   v  \\
\end{matrix} \right]=\mathbf{O}-{{\mathbf{V}}_{\mathbf{0}}}.
\label{18}
\end{equation}
\par This means the barycentric coordinates $(u,v)$ and the distance $s$, from the ray origin to the intersection point can be found by solving the linear system of equation. Denoting ${{\mathbf{E}}_{\mathbf{0}}}=\mathbf{O}-{{\mathbf{V}}_{\mathbf{0}}}$,${{\mathbf{E}}_{\mathbf{1}}}={{\mathbf{V}}_{\mathbf{1}}}-{{\mathbf{V}}_{\mathbf{0}}}$ and ${{\mathbf{E}}_{\mathbf{2}}}={{\mathbf{V}}_{\mathbf{2}}}-{{\mathbf{V}}_{\mathbf{0}}}$, it can be obtained by the Cramer's rule as
\setcounter{equation}{18}
\begin{equation}
\begin{array}{l}
\left[ \begin{matrix}
   s  \\
   u  \\
   v  \\
\end{matrix} \right]=\dfrac{\text{1}}{\left[ \left| \begin{matrix}
   -\mathbf{D} & {{\mathbf{E}}_{\mathbf{1}}} & {{\mathbf{E}}_{\mathbf{2}}}  \\
\end{matrix} \right| \right]}\left[ \begin{matrix}
   \left| \begin{matrix}
   {{\ \mathbf{E}}_{\mathbf{0}}} & {{\mathbf{E}}_{\mathbf{1}}} & {{\mathbf{E}}_{\mathbf{2}}}  \\
\end{matrix} \right|  \\
   \left| \begin{matrix}
   -\mathbf{D} & {{\mathbf{E}}_{\mathbf{0}}} & {{\mathbf{E}}_{\mathbf{2}}}  \\
\end{matrix} \right|  \\
   \left| \begin{matrix}
   -\mathbf{D} & {{\mathbf{E}}_{\mathbf{1}}} & {{\mathbf{E}}_{\mathbf{0}}}  \\
\end{matrix} \right|  \\
\end{matrix} \right] \vspace{3pt}\\
\ \ \ \ \ \ \ =\dfrac{1}{\mathbf{K}\cdot {{\mathbf{E}}_{\mathbf{1}}}}\left[ \begin{matrix}
   \mathbf{Q}\cdot {{\mathbf{E}}_{\mathbf{2}}}  \\
   \mathbf{K}\cdot {{\mathbf{E}}_{\mathbf{0}}}  \\
   \mathbf{Q}\cdot \mathbf{D}  \\
\end{matrix} \right]
\label{19}
\end{array}
\end{equation}
\noindent where $\mathbf{Q}=({{\mathbf{E}}_{\mathbf{0}}}\times {{\mathbf{E}}_{\mathbf{1}}})$ and $\mathbf{K}=(\mathbf{D}\times {{\mathbf{E}}_{\mathbf{2}}})$. By calculating the solution, the location of intersection is obtained to determine whether it is a LoS path. Then, the LoS probability can be found by counting the LoS pairs divided by the total number.
\vfill









\end{document}